\documentstyle[a4wide,12pt]{article}
\title{Hilbert's Incompleteness, Chaitin's $\Omega$ number and
Quantum Physics}
\author{Tien D Kieu~\footnote{email: kieu@swin.edu.au}\\
Centre for Atom Optics and Ultrafast Spectroscopy,\\Swinburne
University of Technology, Hawthorn 3122, Australia}
\begin{document}
\maketitle
\begin{abstract}
To explore the limitation of a class of quantum algorithms originally
proposed for the Hilbert's tenth problem, we consider two further classes of
mathematically non-decidable problems, those of a modified version of the 
Hilbert's tenth problem and of the computation of the Chaitin's 
$\Omega$ number, which is a representation of the G\"odel's 
Incompletness theorem.  Some interesting connection to Quantum Field 
Theory is pointed out.
\end{abstract}

\section*{Introduction}
Quantum physical processes have been considered for the purpose of
computation for some time, and more recently some are proposed in~\cite{kieu}
for certain class of mathematical 
non-computables associated with the Hilbert's tenth problem and, equivalently, 
the Turing 
halting problem~\cite{hilbert}.  There is, however, a hierarchy of non-
computable/undecidable problems~\cite{recursive} of which the Turing
halting problem is at the lowest level.  Another well-known non-
decidable result is the G\"odel's Incompleteness Theorem about the
incompatibility of consistency and completeness of Arithmetics, see~\cite{godel}
for a readable account.  As the mathematical property of being 
Diophantine, which is at the heart of the Turing halting problem, is not 
a sufficient condition for that of being arithmetic~\cite{davis}, which
concerns the G\"odel's, the quantum algorithm cited above has no direct
consequence on the G\"odel's theorem.

Notwithstanding this, in order to explore the limitation of our 
algorithm we consider two further classes of mathematically non-decidable 
problems.  We find some connection between 
Quantum Field Theory and the Chaitin's $\Omega$ number which intimately 
links to the G\"odel's theorem.
Along the way we also consider some modified versions of the Hilbert's
tenth problem where the question about a Diophantine equation is now
concerning the finiteness of the number of its integer solutions.  

In the next section we recall the proposed quantum algorithm, and its physical
principles, for the Hilbert's tenth problem to set the scene.

\section*{Quantum algorithm for the Hilbert's tenth problem}
All the considerations in this work are related through the Diophantine
equation which is a polynomial equation with integer coefficients and
many unknowns,
\begin{eqnarray}
P(x_1,\cdots,x_K) &=& 0,
\label{dio}
\end{eqnarray}
of which it suffices to consider only non-negative integer solutions.
The Hilbert's tenth problem asks for a universal algorithm to determine
whether such an equation has any solution or not.  

Instead of looking for the zeroes of the polynomial in~(\ref{dio}) 
over the non-negative integers, which may not exist, we will search for 
the absolute minimum of its square, $(P(x_1,\cdots,x_K))^2$, which always
exists and finite.  Quantum mechanics facilitates this search through the
ground state of the corresponding hamiltonian
\begin{eqnarray}
H_P &=& \left(P(a_1^\dagger a_1,\cdots, a_K^\dagger a_K)\right)^2,
\label{hamil}
\end{eqnarray}
where $(a, a^\dagger)$ commute for different indices and are the
creation and annihilation operators acting on some Fock space.  The
ground state has the energy which is exactly the searched-for 
absolute minimum. It also carries some information, which can be revealed
through measurements, about the set of some integers $(n_1, \cdots,
n_K)$ at which the minimum is obtained.  In~\cite{kieu} we consider the
processes to simulate the desired hamiltonian, and to obtain the ground
state through some adiabatic evolution starting from some readily
constructed ground state of some initial hamiltonian.  One such initial
hamiltonian is
\begin{eqnarray}
H_I &=& \sum_{i=1}^K (a_i^\dagger -\alpha_i^*)(a_i - \alpha_i),
\nonumber
\label{ihamil}
\end{eqnarray}
where $\alpha_i$ are complex numbers.  This hamiltonian has discrete 
spectrum.  Its ground state is
$\otimes^K_{i=1}|\alpha_i\rangle$, a direct product of the coherent
state $|\alpha_i\rangle$,
\begin{eqnarray}
a_i |\alpha_i\rangle &=& \alpha_i |\alpha_i\rangle,
\nonumber
\label{coher}
\end{eqnarray}
which can be produced by stabilised lasers in the case quantum optical systems 
are used for computation.  Note that all 
the hamiltonians and observables involved have discrete eigenvalues. 
In a generic case, it can be argued~\cite{kieu3} that there should be no 
level crossing in the interpolating time, except at the end point $t = T$
where the adiabatic process ends with the hamiltonian~(\ref{hamil}) which 
has some obvious symmetry.  This finiteness of the gap results in the
finiteness of the running time $T$.

\section*{Another decision problem}
We can also modify the Hilbert's tenth problem and ask a different 
question~\cite{davis} whether the equation~(\ref{dio}) has a finite number 
of non-negative integer solutions (including the case it has no solution) or 
an infinite number.  In general, we cannot tell the degree of degeneracy of the 
ground state, so we will consider the most direct application of the algorithm 
of the last section as follows.

We first note that if the Diophantine equation has only a finite number
of solutions then the shifting in the first argument, say, of the Diophantine
polynomial will render the equation
\begin{eqnarray}
P(x_1 + L,\cdots,x_K) &=& 0
\label{shifted}
\end{eqnarray}
has no more solutions for any $L$ larger than the maximum integer value of 
$x_1$ of all the solutions of the original, unshifted Diophantine
equation.  This trick works because the $x_i$'s are restricted to 
non-negative integer numbers.  With this observation, we could apply the
quantum algorithm of the last section for 
\begin{eqnarray}
P(x_1 + i,\cdots,x_K) &=& 0
\label{shifted2}
\end{eqnarray}
for $i = 0, 1, 2, \cdots$ in turn.  If for some $i=L$ the algorithm
tells us that the equation has no solution we could conclude that it has
only a finite number of solutions.  Otherwise, we will go on forever when 
there is an infinite number of solutions.  This is thus a kind of halting 
problem of different flavour.

To proceed further, we observe that
\begin{eqnarray}
\sum_{i=0}^\infty \left(P(x_1 + i,\cdots,x_K)\right)^2
\label{sum} 
\end{eqnarray}
will have zero absolute minimum value if and only if
equation~(\ref{dio}) has an infinite number of non-negative integer
solutions.

It is clear that the absolute minimum of the sum~(\ref{sum}) diverges 
if the equation has only a finite number of solutions.  To regularise 
the absolute minimum of the sum and make it convergent
we can introduce some coefficients $\beta_i$,
\begin{eqnarray}
\sum_{i=0}^\infty \beta_i \left(P(x_1 + i,\cdots,x_K)\right)^2.
\label{regul}
\end{eqnarray}
For example,
\begin{eqnarray}
\beta_i &=& 1/i!
\label{beta}
\end{eqnarray}
will suffice because, for large $i$, the factorial function increases
much faster than any polynomial
\begin{eqnarray}
\min \left(P(x_1 + i, x_2, \cdots,x_K)\right)^2 \le \left(P(i, 0, \cdots, 0)\right)^2 << i!.
\label{min}
\end{eqnarray}
The first inequality follows from the fact that the point 
$(i,0,\cdots,0)$ belongs to
the boundary of the domain of the square of the Diophantine polynomial.

We could now apply the quantum algorithm of the last section to form the
hamiltonian
\begin{eqnarray}
H_P &=& \sum_{i=0}^\infty \beta_i \left(P(a^\dagger_1 a_1 + i,\cdots, 
a^\dagger_K a_K)\right)^2
\label{hamil2}
\end{eqnarray}
and try to obtain its ground state as before.  As with regularisation in
Quantum Field Theory, we eventually remove the regulator through some
dependence on some parameter $0<s<1$
\begin{eqnarray}
\lim_{s\to 0} \beta_i(s) &=& 1,
\label{regul}
\end{eqnarray}
for instance, $\beta_i(s) = (1/i!)^s$.  If we could tell that the ground
state energy is zero as 
$s\to 0$, starting from $\beta_i(1)$, throughout then the equation has an 
infinite number of solutions.  However, implementing such limit may not
be achievable since the coefficients on the right hand side 
of~(\ref{hamil2}), unlike before, are no longer integers
but a convergent series for each of them.  We may not be able 
in general to evaluate these series, even though they are made convergent, 
in order to simulate the hamiltonian to the desired precision.
(There should be no problem of the unbounded amount of energy that would be required in the 
regulator-removal limit, $s\to 0$ if it could be implemented,
because as soon as the ground state energy is distinguishable from
zero then we know that the equation has a finite number of solutions.)

\section*{God\"el's Incompleteness and Chaitin's $\Omega$ number}
The G\"odel's Second Incompleteness Theorem~\cite{godel} is the negative 
answer for
the Hilbert's second problem about formalisation of mathematics.
G\"odel's theorem stipulates that Arithmetics cannot be both consistent
and complete.  Interpretation of this theorem tells us that if a formal
system, with a finite set of axioms and inference rules, is consistent 
and could include Arithmetics then there exist unprovable statements 
within that system; incompleteness thus entails.  (Interestingly, it may be 
precisely the statement about consistency of the system that is 
unprovable.)  Our quantum algorithm for the Hilbert's tenth problem only
deals with the mathematical property of being Diophantine, which is a
sub-property of arithmetic~\cite{davis}, and thus cannot resolve the 
G\"odel's decision problem which concerns with the property of being 
arithmetic in general.

Chaitin has approached this problem from the perspectives of Algorithmic
Information Theory~\cite{chaitin2, chaitin3} and shown that there exist many
unprovable statements in Arithmetics simply because they have irreducible 
algorithmic contents, measurable in bits, that are more than the complexity, 
also measurable in bits, of the finite set of axioms and inference rules 
for the system.  That is, there is randomness even in pure mathematics.  
And more frustratingly, we can never prove randomness since we could 
only ever deal with finite axiomatic complexity.

Chaitin, to illustrate the point, has introduced the number $\Omega$ as
the halting probability for a random program, with some random input, 
being emulated by a particular Turing machine.
It is also an average measure over all programs run on the universal Turing 
machine.
This number has many interesting properties, and has been generalised to a 
quantum version~\cite{svozil}, but we only make 
use here the linkage between this number and polynomial Diophantine 
equations.  When expressed in binary, the value of the $k$-th bit of 
$\Omega$ is 0 or 1 depending on some Diophantine equation 
corresponding to the Turing machine in consideration,
\begin{eqnarray}
C(k,N,x_1,\cdots,x_K) &=& 0.
\label{omeg}
\end{eqnarray}
For a given $k>0$, the $k$-th bit is determined by whether there are 
finitely or infinitely many values for the parameter $N>0$ for which 
the equation above has solutions in non-negative integers 
$(x_1, ..., x_K)$.

We now try to adapt our quantum algorithm for the Hilbert's tenth
problem for this kind of equations which represent the G\"odel's 
theorem in a different guise.  But then we could only come up with the 
hamiltonian
\begin{eqnarray}
H_P &=& \sum_{N=0}^\infty \beta_N(s) \left(
C(k,N,a_{1+NK}^\dagger a_{1+NK},\cdots,a_{(1+N)K}^\dagger
a_{(1+N)K})\right)^2,
\label{qft}
\end{eqnarray}
where we have appealed to the framework of Quantum Field Theory for the
appearance of a countably infinite number of pairs of operators $(a,
a^\dagger)$.  Theoretically, the ground state of~(\ref{qft}) has 
relatively zero energy in the regulator-removal limit, $s\to 0$, if 
and only if the equation~(\ref{omeg}) has solutions for an infinitely many 
values for the parameter $N$.

The connection to Quantum Field Theory is interesting but it makes the
whole exercise somewhat academic because we yet know how to create QFT 
hamiltonians on demand.

\section*{Concluding remarks}
If, as we hope, the quantum algorithm for the Hilbert's tenth problem 
can be implemented then all is the better.  However, there might be some
fundamental physical principles, not those of practicality, which prohibit
the implementation.  Or, there might be not enough physical resources 
(ultimately limited by the total energy and the lifetime of the universe) 
to satisfy the execution of the algorithm.  
In either cases, the exercise is still very interesting as 
the unsolvability of those problems and the limit of mathematics
itself are also dictated by physical principles and/or resources.

Generalised noncomputability and undecidability set the boundary for computation 
carried out by
mechanical (including quantum mechanical) processes, and in 
doing so help us to understand much better what can be so computed.
With this in mind, we consider some adaptations of our quantum algorithm 
for the Hilbert's tenth problem to some modified version of 
the Hilbert's tenth problem.  We also consider the 
computation of Chaitin's $\Omega$ number.  These problems are all inter-related through 
questions about existence of solutions of Diophantine equations.
Among the implementation difficulties, some interesting connections to Quantum
Field Theory are pointed out along the way but we do not know how to create QFT hamiltonians on demand.

\section*{Acknowledgements}  I am indebted to Alan Head for discussions, 
comments and suggestions; and would like to thank Cristian Calude for
informing me the recent works on the $\Omega$ and quantum $\Omega$
numbers.

\end{document}